\documentclass[a4paper,twocolumn,%tightenlines,
english,aps,pre,floatfix,groupedaddress,showpacs,nofootinbib]{revtex4-1}
\usepackage[T1]{fontenc}
\usepackage[latin1]{inputenc}
\usepackage{amsmath}
\usepackage{babel}
\usepackage{graphics}
\usepackage{amssymb}
\usepackage{mathrsfs}
\usepackage{dcolumn}


\begin{document}
\title
{Current-activity versus local-current fluctuations in driven flow with exclusion}
\author {S. L. A. \surname{de Queiroz}}
\email{sldq@if.ufrj.br}
\affiliation{Instituto de F\'\i sica, Universidade Federal do
Rio de Janeiro, Caixa Postal 68528, 21941-972
Rio de Janeiro RJ, Brazil}

\date{\today}

\begin{abstract} 
We consider fluctuations of steady-state current activity, and of its dynamic counterpart,
the local current, for the one-dimensional totally asymmetric simple exclusion 
process. The cumulants of the integrated activity behave similarly 
to those of the local current, except that they do not capture the anomalous scaling behavior
in the maximal-current phase and at its boundaries. 
This indicates that the systemwide sampling at equal times, characteristic of the instantaneous
activity, overshadows the subtler effects which come about from non-equal time correlations,
and are responsible for anomalous scaling. We show that apparently conflicting results
concerning asymmetry (skewness) of the corresponding distributions can in fact be reconciled,
and that (apart from  a few well-understood exceptional cases) 
for both activity and local current one has positive skew deep within the low-current
phase, and negative skew everywhere else.
\end{abstract}
%05.40.-a Fluctuation phenomena, random processes, noise, and Brownian motion
%02.50.-r Probability theory, stochastic processes, and statistics
%05.70.Fh Phase transitions: general studies
\maketitle
%\tightenlines
 
\section{Introduction} 
\label{intro} 
In this paper we consider fluctuations of the steady-state current, and of its close 
relative, the current {\em activity}, in the one-dimensional totally asymmetric simple 
exclusion process (TASEP). This model is among the simplest in non-equilibrium physics, 
while at the same time exhibiting many non-trivial 
properties~\cite{derr98,sch00,derr93,rbs01,be07,cmz11}. Some relevant 
developments in the study of current fluctuations are as follows:
exact expressions for the diffusion constant were found 
for systems with periodic (PBC)~\cite{dem93} and open~\cite{dem95} 
boundary conditions (BC);
the full probability distribution function (PDF) of current fluctuations was
similarly considered for both PBC~\cite{dl98} and open~\cite{bd06} BC. 
Very recently, a number
of new results have been found for current fluctuations in systems with open 
BC~\cite{kvo10,gv11,lm11,ess11}. 
 
In a recent publication~\cite{rbsdq12}, exact and numerical results were given for
steady-state current activity fluctuations in the 
one-dimensional TASEP, for both periodic and open boundary conditions. 
By making use of the known steady-state 
operator algebra~\cite{derr93}, exact expressions were derived for the three
lowest moments of the activity PDF, which fully display their finite-size
dependence. All these were confirmed to excellent degree of accuracy by numerical 
simulations. The results of Ref.~\onlinecite{rbsdq12} extend and complement earlier 
analytic work on the joint distribution of current activity and density 
for the TASEP~\cite{ds04,ds05}. We recall that the current activity  
(henceforth denominated simply {\em activity}) is not identical to 
the standard current, although the first moments of the respective distributions
coincide. As explained below, the former quantity is {\em static},
in this sense akin to the instantaneous (local or global) particle density,
while the latter  is a dynamic one.

Many exact results available for current fluctuations pertain to the 
infinite-system limit~\cite{dl98,bd06,ess11}, although the diffusion
constant has been calculated for finite systems~\cite{dem93,dem95}. Finite-size 
effects have been considered also, e.g., in Refs.~\onlinecite{kvo10,gv11,lm11}.   

Our main purpose here is to exploit the possible connections between activity- and current
fluctuations. Given that the former quantity has proved amenable to such detailed 
description, it is desirable to check whether its properties can help explain 
any relevant aspects of the latter. 

In the time  evolution  of the $1+1$ dimensional TASEP,
the particle number $n_\ell$ at lattice site $\ell$ can be $0$ or $1$, 
and the forward hopping of particles is only to an empty adjacent site. 
The stochastic character comes from random selection of site occupation 
update~\cite{rsss98,dqrbs08}: if site $\ell$ is chosen for update,
the instantaneous current across the bond from $\ell$ to $\ell +1$ is
given by  $J_{\ell,\ell+1}= n_\ell (1-n_{\ell+1})$~. 

With open boundary conditions, the case  considered here,
the additional externally-imposed parameters are: 
the injection (attempt) rate $\alpha$ at the left end,
and the ejection rate $\beta$ at the right one.
The phase diagram in $\alpha$--$\beta$ parameter space, 
reproduced in Figure~\ref{fig:obc_pd} below,
is known exactly, as
well as many other steady state 
properties~\cite{derr98,sch00,derr93,rbs01,be07,ds04,ds05,nas02,ess05}.

\begin{figure}
{\centering \resizebox*{3.3in}{!}{\includegraphics*{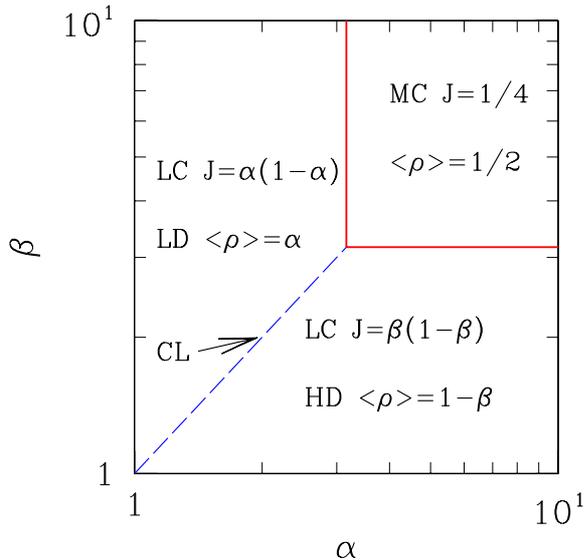}}}
\caption{(Color online) 
Phase diagram of TASEP with open BC. Values of steady-state current $J$
and density $\langle \rho \rangle$ correspond to $L \to \infty$. The
low-current (LC), low-density (LD) phase is separated from the LC, high-density
(HD) phase by a first-order transition 
%%%%%%%%%%% Changed by editors' request %%%%%%%%%%%%%%%%%%%%%%%%%%%%%%%
along the coexistence line CL (long-dashed, blue). The 
maximal-current (MC) phase is separated from the LC phases by second-order
transition lines (full red lines).
%%%%%%%%%%%%%%%%%%%%%%%%%%%%%%%%%%%%%%%%%%%%%%%%%%%%%%%%%%%%%%%%%%%%5
} 
\label{fig:obc_pd}
\end{figure}

The total (instantaneous) activity $A$ within the system is defined as the number 
of bonds that can facilitate a transition of a particle in the immediate
future. Thus it equals the number of pairs of neighboring sites that 
have a particle to the left and a hole to the right~\cite{ds04,ds05}.
For systems with open BC, one usually includes 
also the injection  and ejection bonds at the system's ends, though these have to be 
weighted by the respective injection and ejection rates, $\alpha$ and $\beta$.

For an $L$--site system with open BC 
($L$ sites and $L+1$ bonds, including the injection and ejection ones), one has:
\begin{equation}
A =\alpha\,(1-n_1)+ \sum_{\ell=1}^{L-1} n_\ell\,(1-n_{\ell+1})+\beta\,n_L\quad
{\rm (Open\ BC)}\ .
\label{eq:adef}
\end{equation}

The activity is a snapshot of the system
at a given moment in its evolution; in this sense, it is as much of a static quantity 
as, for example, the instantaneous global density. By contrast, the current is a
dynamic object, as it reflects the stochastically-determined  particle 
displacements which actually take place during a unit time interval.  
The investigation of current fluctuations is usually carried out
by examining the total charge (i.e., number of particles) crossing a given bond, 
during a long time interval in the steady state regime~\cite{dem95,dl98,bd06,kvo10,gv11,lm11,ess11}.

The (properly normalized) first moments of activity and current PDFs coincide.
For $A$ as defined in Eq.~(\ref{eq:adef}) one has:
\begin{equation}
J=\frac{1}{L+1}\,\langle A\rangle\qquad{\rm (Open\ BC)}\ ,
\label{eq:a-j}
\end{equation}
where $J$ is the average steady-state current through any bond in the system,
and brackets $\langle \dots\rangle$ denote ensemble averages. 
The above equality can be understood by recalling that
successive steady-state snapshots (activity configurations) are generated via 
the intervening particle hoppings, which constitute realizations of the 
system's current. In our simulations of Ref.~\onlinecite{rbsdq12} we verified that this 
property holds, to within numerical accuracy, in all cases investigated there.
However, the connection at this level 
is not sufficient to warrant equality of higher moments of the PDFs. 

Here we restrict ourselves to the second and third cumulants of current- and 
activity PDFs. These already 
provide significant illustrations of the diversity of behavior which it is our
purpose to investigate. While more precise definitions are deferred to 
Section~\ref{sec:th} below, Tables~\ref{t1} and~\ref{t2}, and Figure~\ref{fig:pd}, 
give a broad perspective of properties, such as system-size dependence or $L \to
\infty$ limiting values, 
of these cumulants, or quantities associated to them (respectively, variance and 
skewness~\cite{rbsdq12,numrec} for  activity; diffusion constant~\cite{dem95} and 
skew~\cite{lm11,ess11} for current). All quoted results for activity statistics
were previously published in Ref.~\onlinecite{rbsdq12}, having been
obtained analytically, and supported by numerical simulations; 
those for current statistics are exact predictions found by assorted analytic 
techniques~\cite{dem95,lm11,ess11}.

\begin{table}
\caption{\label{t1} 
For systems with $L$ sites, $\alpha$, $\beta$ as specified, and
$\gamma \equiv {\rm min} (\alpha,\beta)$, 
system-size dependence of: (i)  variance for (normalized) activity PDF
[$\,$see Eq.~(\protect{\ref{eq:a-j}})$\,$], 
and (ii) diffusion constant~\cite{dem95} for current (in the latter, values quoted  
for LC phase correspond to $L \to \infty$).
}
\vskip 0.2cm
\begin{ruledtabular}
\begin{tabular}{@{}ccc}
\multicolumn{3}{c}{Variance / Diffusion constant}\\
\hline\noalign{\smallskip}
Region &  Activity~\cite{rbsdq12}   & Current \\
\hline\noalign{\smallskip}
LC $[\,\gamma <1/2\,]$ &   $L^{-1}$  &  
$\gamma(1-\gamma)(1-2\gamma)$~\cite{dem95,lm11,ess11} \\
MC  $[\,(\alpha,\beta)> 1/2\,$]  & $L^{-1}$  & 
$L^{-1/2}$~\cite{dem95,lm11} \\
\hline\noalign{\smallskip}
\end{tabular}
\end{ruledtabular}
\end{table}

\begin{figure}
{\centering \resizebox*{3.3in}{!}{\includegraphics*{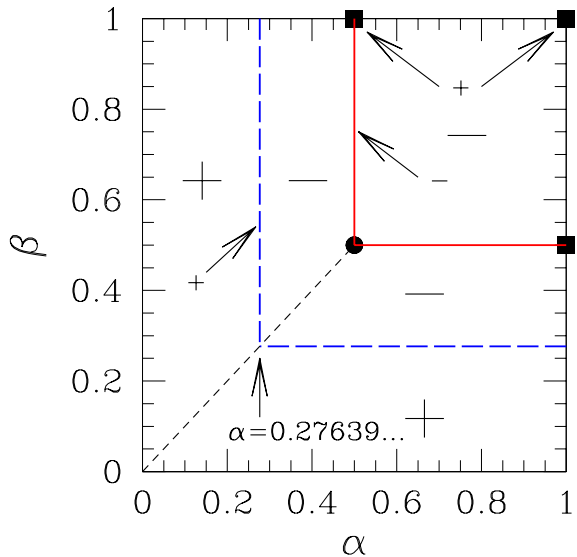}}}
\caption{(Color online)
The sign of skewness $S$~\cite{rbsdq12,numrec} of the activity PDF in the various 
regions of the  $\alpha-\beta$ phase diagram is shown.
The large--$L$ dependence, $|S| \propto L^{-x}$, is $x=1/2$
in the LC phases [${\rm min}\,(\alpha,\beta) < 1/2$] 
except for the long-dashed (blue) lines at ${\rm min}\,(\alpha, \beta)=
\frac{1}{2}-\frac{\sqrt{5}}{10}=0.27639 \dots$ $[\,x=3/2\,]$.
In the MC phase $\alpha, \beta>1/2$, and on the full (red) lines 
separating MC and LC 
phases: $x=3/2$. Full squares: $x=5/2$.
On the $\alpha=1$ and $\beta=1$ lines, $S$
has the same sign and $L$-dependence as in the respective adjacent regions,
except for the points marked by full squares.
The circle marks $(\alpha,\beta)=(1/2,1/2)$ where $S \equiv 0$.
The short-dashed line is the coexistence line between HD and LD phases.
Because of particle-hole duality, $S$ is the same for pairs of points
symmetric with respect to  the $\alpha=\beta$ line. (Adapted from Figure 8 of 
Ref.~\onlinecite{rbsdq12}).}
\label{fig:pd}
\end{figure}

\begin{table}
\caption{\label{t2} 
For systems with $\alpha$, $\beta$ as specified,  and 
$\gamma \equiv {\rm min} (\alpha,\beta)$, predicted 
$L \to \infty$ properties of skew (third cumulant) of current PDF.}
\vskip 0.2cm
\begin{ruledtabular}
\begin{tabular}{@{}cc}
\multicolumn{2}{c}{Skew}\\
\hline\noalign{\smallskip}
Region/point & Analytic prediction \\
\hline\noalign{\smallskip}
LC $[\,\gamma <1/2\,]$ & $\gamma(1-\gamma)(1-6\gamma+
6\gamma^2)$~\cite{lm11,ess11}\\
MC  $[\,(\alpha,\beta)> 1/2\,$] & $<0$~\cite{lm11} \\
$(\alpha,\beta)=1$ & $-0.009\,0978\,\dots$~\cite{lm11}\\ 
\hline\noalign{\smallskip}
\end{tabular}
\end{ruledtabular}
\end{table}
  
So, all quantities associated with activity statistics vanish as $L \to \infty$
(except for the skewness  at $\alpha=\beta=1/2$ which vanishes 
identically~\cite{rbsdq12}, see caption to Fig.~\ref{fig:pd}),
while those related to the standard current usually
approach finite limits (except for the diffusion constant in the MC phase,
see Table~\ref{t1}, and the skew on the lines  ${\rm min}(\alpha,\beta)
\equiv \gamma_0=\frac{1}{2}-\frac{\sqrt{5}}{10}$ deep inside the LC phase, see 
Table~\ref{t2}).

Here we wish to pin down the causes for such variety of behavior in apparently
similar fluctuation-related quantities. As shown in the following, a
fruitful line of enquiry is to probe the relative
importance of different types of correlations which occur in this context:
local versus global (i.e. systemwide), as well as equal- versus non-equal time
(see Section~\ref{sec:th} for precise definitions).

Section~\ref{sec:th} below recalls selected existing results, and
gives a theoretical background to the concepts
used in this work. In Section~\ref{sec:nr}, our numerical simulations are described,
and their corresponding results are exhibited. 
In Sec.~\ref{sec:conc} we provide a global analysis of the numerical results; 
finally, concluding remarks are made.

\section{Theory}
\label{sec:th}

In studies of current fluctuations for the TASEP, one considers the
steady-state current through a specified bond connecting sites $\ell$ and $\ell+1$,
henceforth denoted by $J_\ell$, and the
associated integrated charge ${\widetilde Q}_\ell(t) \equiv \int_0^t J_\ell (t^\prime)\,dt^\prime$.
With open BC, the leftmost (injection) bond $\ell=0$ is usually singled out
for examination, 
although the results for the PDF of steady-state current
fluctuations would be the same for any choice of $\ell$~\cite{dem95}.
Thus, here we frequently omit bond labels wherever this does not give rise to
ambiguity.  
Also, it is usual to remove the linear term from the integrated charge, and
to consider instead:
\begin{equation}
Q(t) \equiv {\widetilde Q}(t)-Jt\ ,
\label{eq:qtilde}
\end{equation}
so $\langle Q(t) \rangle \equiv 0$. Closed-form expressions are available
for $J$ as a function of $\alpha$, $\beta$, and number of sites $L$~\cite{derr93}; see,
e.g., Eqs.~(29) and~(32)--(34) of Ref.~\onlinecite{rbsdq12}. 

For $t \to \infty$, exact expressions for the lowest-order cumulants, $\langle\langle Q^n
\rangle \rangle$, of the integrated current have been proposed: for $n=2$ and any $L$, 
everywhere on 
the phase diagram~\cite{dem95}; for $n=1-3$ and $L \to \infty$, anywhere except for 
the maximal-current phase $\alpha$, $\beta \geq 1/2$~\cite{ess11}; and 
(via a parametric representation) for any $\alpha$, $\beta$, and $L$, essentially for all 
$n$~\cite{lm11}. In the latter, explicit expressions are forthcoming only for $L \to \infty$ away 
from the maximal-current phase, and for $n \leq 3$, any $L$, at $(\alpha,\beta)=(1,1)$.
The above results agree with each other, wherever comparison is possible. Furthermore,
on the basis of general properties of the associated generating function (see e.g. 
Ref.~\onlinecite{lm11}), the assumption is made that all cumulants scale linearly with time, 
so the quantities of interest are the $E_n \equiv \langle\langle Q^n\rangle \rangle/t$,
which are frequently referred to as {\em cumulants of the local current}~\cite{ess11}. 
So, $E_2$ is the diffusion constant, and $E_3$ the skew, both mentioned in connection with Tables~\ref{t1} and~\ref{t2}.
 
A direct elementary illustration of such linear dependence can be given by
considering the variance $\langle\langle {Q}^2(t)\rangle\rangle$~. 
With $\delta J_\ell(t) \equiv J_\ell(t)-J$, assuming an exponential decay of the
current-current correlation function, $\langle \delta J_\ell(t^\prime)\,
\delta J_\ell(t^{\prime\prime})\rangle
\sim e^{-|t^\prime-t^{\prime\prime}|/\tau}$, and using steady-state properties, 
one has:
\begin{eqnarray}
\langle\langle {Q}^2(t)\rangle\rangle =\int_0^t dt^\prime\int_0^t dt^{\prime\prime}
\langle \delta J_\ell(t^\prime)\,\delta J_\ell(t^{\prime\prime})\rangle=\nonumber\\
\quad ({\rm for}\ t \gg \tau)=t\,\int_0^t ds\,\langle \delta J_\ell(0)\,
\delta J_\ell(s)\rangle= t\tau\ .
\label{eq:q2proptot}
\end{eqnarray}
An interesting exception to the above has been pointed out in 
Ref.~\onlinecite{kvo10}, where theoretical and numerical arguments are presented to show that
 $\langle\langle Q^n\rangle \rangle \sim t^{\,n/3}$ for $n>1$
at the second-order phase boundary between low- and maximal-current phases. We shall return to
this later.

As recalled in Eq.~(\ref{eq:q2proptot}), the cumulants 
$\langle\langle Q^n \rangle \rangle$, $n>1$, involve unequal-time 
correlations~\cite{dem95}, because the integrated charge accumulates local current
fluctuations over time. In contrast, see Eq.~(\ref{eq:adef}),
the nontrivial features of (instantaneous) activity statistics
arise because it adds contributions from all sites in the system, thus it
depends on non-local, equal-time, correlations~\cite{rbsdq12}. In fact, it is
because  $A(t)$ is the sum of ${\mathcal O}(L)$ equal-time stochastic (albeit not independent) 
variables that the variance 
and skewness (and, most likely, higher-order moments) of its PDF
always approach zero for large $L$~\cite{rbsdq12}. 
The spatially local character of the $\langle\langle Q^n \rangle \rangle$,
in turn, implies less severe $L$-dependent effects on these quantities; thus the $E_n$
generally converge towards non-zero values for $L \to \infty$~\cite{dem95,ess11,lm11}.   

As stated in Section~\ref{intro}, 
we wish to consider the relative importance of 
local and non-local  (equal and non-equal time) correlations.   
We then define a hybrid quantity, the {\em position-averaged} (instantaneous) {\em current}
${\mathcal J}(t)$:
\begin{equation}
{\mathcal J}(t)=\frac{1}{L+1}\sum_{\ell=0}^L J_\ell(t)\ .
\label{eq:jint}
\end{equation}
The ensemble average of ${\mathcal J}(t)$ coincides with that of the normalized activity, see 
Eqs.~(\ref{eq:adef}) and~(\ref{eq:a-j}), and of course with the steady-state current $J$.
The cumulants of its integral over time, ${\mathscr Q}(t) \equiv \int_0^t{\mathcal 
J}(t^\prime)\,
dt^\prime$ incorporate both equal- and unequal-time (as well as local and non-local) 
correlations. For consistency, see Eq.~(\ref{eq:qtilde}),
we also subtract the linear contribution, $Jt$, from ${\mathscr Q}(t)$.

In order to draw relevant distinctions and similarities with ${\mathscr Q}(t)$, we
also consider the integrated 
(normalized) activity, ${\mathcal A}(t) \equiv(L+1)^{-1}\,\int_0^t A(t^\prime)\,dt^\prime$
(again, with the linear contribution $Jt$ subtracted).

In what follows, we show results of numerical simulations of ${\mathcal J}(t)$ and 
associated quantities such as the cumulants of ${\mathscr Q}(t)$, as well
as those of ${\mathcal A}(t)$, in comparison with those corresponding to the local current.

\section{Numerical results}
\label{sec:nr}

\subsection{Introduction}
\label{sec:nr-intro}

To make contact with previous work, we usually considered lattices either
with $L=600$ sites, as done in local-current simulations~\cite{kvo10}, or with $L=256$,
for which many results on activity statistics are available~\cite{rbsdq12}. 
It is important to take $L \gg 1$ not only in order to
minimize finite-size effects, but also because some features such as the non-linear
scaling of current cumulants with time, referred to above, occur only during
time "windows" whose width increases as $L$ grows large~\cite{kvo10}.  

In our simulations, a time step is defined as a set of $L$ sequential update attempts,
each of these according to the following rules: (1) select a site at random;
(2a) if the chosen site is the rightmost one and is occupied, then
(3a) eject the particle from it with probability $\beta$; alternatively,
(2b) if the site is the leftmost one and is empty, then (3b) inject a
particle onto it with probability $\alpha$; finally, if neither (2a) nor (2b)
is true, (2c) if the site
is occupied and its neighbor to the right is empty, then (3c) move the particle. 

Thus, in the
course of one time step, some sites may be selected more than once for examination,
and some may not be examined at all. 
This corresponds to the random-sequential update procedure of 
Ref.~\onlinecite{rsss98}.
%%%%%%%%%%%%%%%% Referee's 2nd report, point (1) %%%%%%%%%%%%%%%%%%%%%%%%%
Note that other types of update are possible (e.g., ordered-sequential or
parallel). Though the resulting phase diagrams are similar in all cases
(but not identical: even the average current differs in either  
case, see  Table 1 in Ref.~\onlinecite{rsss98}), 
the updating algorithm which corresponds to the operator algebra 
described in Ref.~\onlinecite{derr93} [$\,$and thus to many
subsequent results either directly derived from that 
algebra~\cite{rbsdq12}, or from Bethe-ansatz techniques which are based
on it~\cite{ess11}$\,$] is random-sequential~\cite{rsss98}.

%%%%%%%%%%%%%%%%%%%%%%%%%%%%%%%%%%%%%%%%%%%%%%%%%%%%%%%%%%%%%%%%%%%%%%%%

For the various sets of $\alpha$, $\beta$ considered here [$\,$because of particle-hole
duality, we take only $\alpha \leq \beta\,$], we found that for $L=600$
and starting from an initial random configuration of occupied sites, one
needs  $n_{\rm in}=10,000$ time steps for steady-state flow to be fully 
established, so that the $\langle\langle Q^n\rangle\rangle$ are free from 
startup effects (for $\alpha=\beta=1$ this remark needs to be qualified, see
 Section~\ref{sec:ab1}). Hence we discarded 
the first $n_{\rm in}$ configurations
when evaluating all quantities discussed in the following. As seen below in 
Figures~\ref{fig:a2qbar},~\ref{fig:q2q2bar}, and~\ref{fig:aq2qbarab1}, 
the characteristic decay times
for the $\langle\langle {\mathscr Q}^n\rangle\rangle$, $\langle\langle {\mathcal 
A}^n\rangle\rangle$
are about twice that, so the respective curves still show some crossover towards pure 
power-law behavior for the first $\sim 10^4$ time steps depicted.  

\subsection{$\alpha+\beta=1$}
\label{sec:a+b1}
%%%%%%%%%%%%%%%%%%%%%%%%Referee's 2nd report, point (2) %%%%%%%%%%%%%%%%%%%%%%%
It is known~\cite{derr93} that, for $\alpha+\beta=1$ the correlations between the 
operators representing particle and vacancy vanish, and they can be represented by 
$c$-numbers. One  consequence of this is that the average current
does not exhibit finite-size effects: $J=\alpha\beta$ for any system size.
Further simplifications occur, so that simple expressions can also be found for 
higher-order moments of the activity
distribution~\cite{rbsdq12} 
[$\,$which is not generally true for points elsewhere on the phase
diagram, with the notable exception of $(\alpha,\beta)=(1,1)$ deep inside the
MC phase, see Section~\ref{sec:ab1}$\,$]. 
Along $\alpha+\beta=1$ we take one point, $(0.3,0.7)$, representative of 
behavior deep inside the LC phase, and another, $(0.5,0.5)$ on the second-order
transition line between LC and MC phases.
%%%%%%%%%%%%%%%%%%%%%%%%%%%%%%%%%%%%%%%%%%%%%%%%%%%%%%%%%%%%%%%%%%%%%%%%%%%%%%

\begin{figure}
{\centering \resizebox*{3.3in}{!}{\includegraphics*{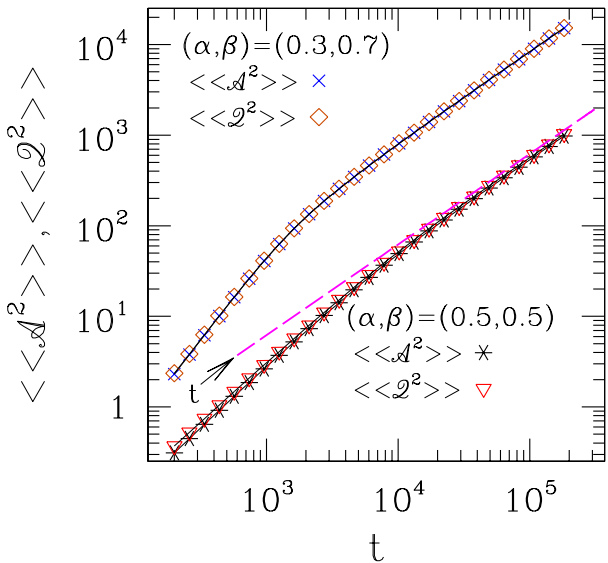}}}
\caption{(Color online) 
Double-logarithmic plot of variances $\langle\langle {\mathcal A}^2\rangle\rangle$, for
the integrated activity ${\mathcal A}(t)$, 
and $\langle\langle {\mathscr Q}^{\,2}\rangle\rangle$ for ${\mathscr Q}(t)$, both
defined below Eq.~(\protect{\ref{eq:jint}}), against time. 
The dashed line is $\propto t$. System size $L=600$. 
} 
\label{fig:a2qbar}
\end{figure}

In Figure~\ref{fig:a2qbar} we present the evolution of 
$\langle\langle {\mathcal A}^2 \rangle\rangle$ and 
$\langle\langle {\mathscr Q}^2 \rangle\rangle$ against time, for $(\alpha,\beta)=(0.3,0.7)$ 
and $(0.5,0.5)$. In both cases the variance initially increases
faster than linearly with time, but eventually  settles to a linear increase.
Also, both quantities
behave identically for $\alpha=0.3$, $\beta=0.7$, while for the special case $\alpha=\beta=0.5$
they follow very close trends,
the ratio $\langle\langle {\mathcal A}^2 \rangle\rangle/\langle\langle {\mathscr Q}^2 
\rangle\rangle$
starting at around $0.85$ and approaching unity as $t$ grows (near the upper limit in the Figure,
it has reached $0.95$). 

One sees that the dominant features of the second-order cumulants of  $\mathscr Q$
essentially coincide with those of $\mathcal A$. As shown below, the
quantitative description of  $n=3$ cumulants, and of their associate skewness properties,
always points to their absolute value being rather small; at the level of
accuracy pursued here the main issue therefore concerns their sign, which 
generally proves to be a robust feature (i.e., independent of whether $\mathscr Q$ or
 $\mathcal A$ is the quantity under consideration).   
Thus, in the following, we usually
display only results for the  $\langle\langle {\mathscr Q}^{\,n}\rangle\rangle$, for comparison 
with the conventional $\langle\langle Q^{\,n}\rangle\rangle$~. 
 
\begin{figure}
{\centering \resizebox*{3.3in}{!}{\includegraphics*{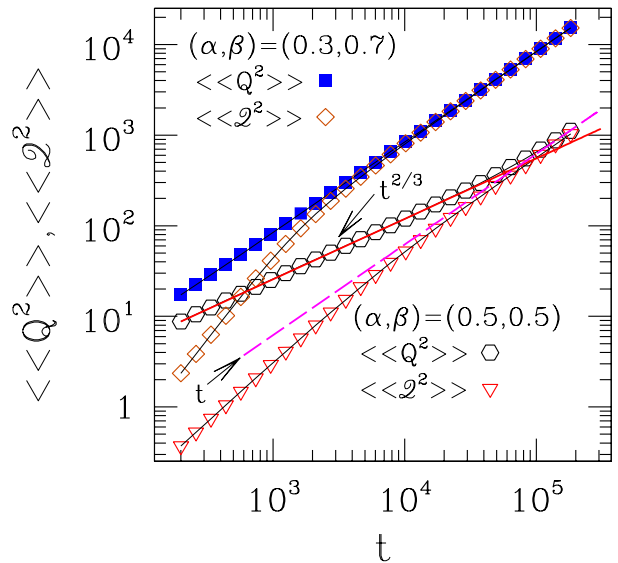}}}
\caption{(Color online) 
Double-logarithmic plot of variances $\langle\langle {Q}^2\rangle\rangle$, for
$Q(t)$ defined in Eq.~(\protect{\ref{eq:qtilde}}), 
and $\langle\langle {\mathscr Q}^{\,2}\rangle\rangle$ for ${\mathscr Q}(t)$
defined below Eq.~(\protect{\ref{eq:jint}}), against time. 
The full line is $\propto t^{2/3}$; the dashed line is $\propto t$. System size $L=600$. 
} 
\label{fig:q2q2bar}
\end{figure}

In Figure~\ref{fig:q2q2bar} we show the cumulants $\langle\langle {Q}^2\rangle\rangle$ and
$\langle\langle {\mathscr Q}^{\,2}\rangle\rangle$ against $t$, for $\alpha=0.3$, $\beta=0.7$,
and for $\alpha=\beta=0.5$~. 

Regarding data for $\langle\langle {Q}^2\rangle\rangle$ in Fig.~\ref{fig:q2q2bar},
for $\alpha=0.3$, $\beta=0.7$ the expected linear
behavior against $t$ is present for the whole time interval shown. On the other hand,
for $\alpha=\beta=0.5$ one sees the anomalous scaling with
$t^{\,2/3}$, referred to above, over a wide time window up to $t \approx 3\times 10^4$. 
Both types of power-law behavior have been reported in
Ref.~\onlinecite{kvo10}, for the same two sets of values of $(\alpha,\beta)$.
For the $\alpha=\beta=0.5$ data, at longer times there is a crossover back 
towards linear scaling. 

Anomalous scaling at $\alpha=\beta=0.5$ has been explained in detail in 
Ref.~\onlinecite{kvo10}. It can be understood by recalling that this
point is on the second-order phase boundary separating low- and 
maximal-current phases
(see Figure~\ref{fig:obc_pd}).
The associated diverging correlation length $\xi$ brings about a diverging relaxation
time $\tau \sim \xi^z$ (where $z=3/2$ is the Kardar-Parisi-Zhang exponent known
to describe TASEP dynamics~\cite{derr98,sch00,rbs01}) which, in turn,
governs the current-current correlation fluctuations and related quantities. 
For systems of finite size $L$, this regime holds only for $t \lesssim L^z$, 
i.e., the "windows" referred to earlier. 
In connection with Eq.~(\ref{eq:q2proptot}), it can be seen that in this case,
the assumption of $t \gg \tau$ fails over the window of anomalous scaling, but
is then restored at longer times; hence, the observed 
crossover towards linear behavior.        

Going now to data for $\langle\langle {\mathscr Q}^2\rangle\rangle$, 
one sees that they eventually merge with the respective $\langle\langle {Q}^2\rangle\rangle$ 
curves. For  $\alpha=\beta=0.5$ such merging happens, of course, at later times
than the window of anomalous scaling. 
It was shown in Ref.~\onlinecite{rbsdq12} that the activity PDF is a pure Gaussian at 
this point; here, as already remarked in connection with Figure~\ref{fig:a2qbar},
we see  that $\langle\langle {\mathscr Q}^2\rangle\rangle$ 
does not capture the anomalous scaling behavior there, either. 

\begin{figure}
{\centering \resizebox*{3.3in}{!}{\includegraphics*{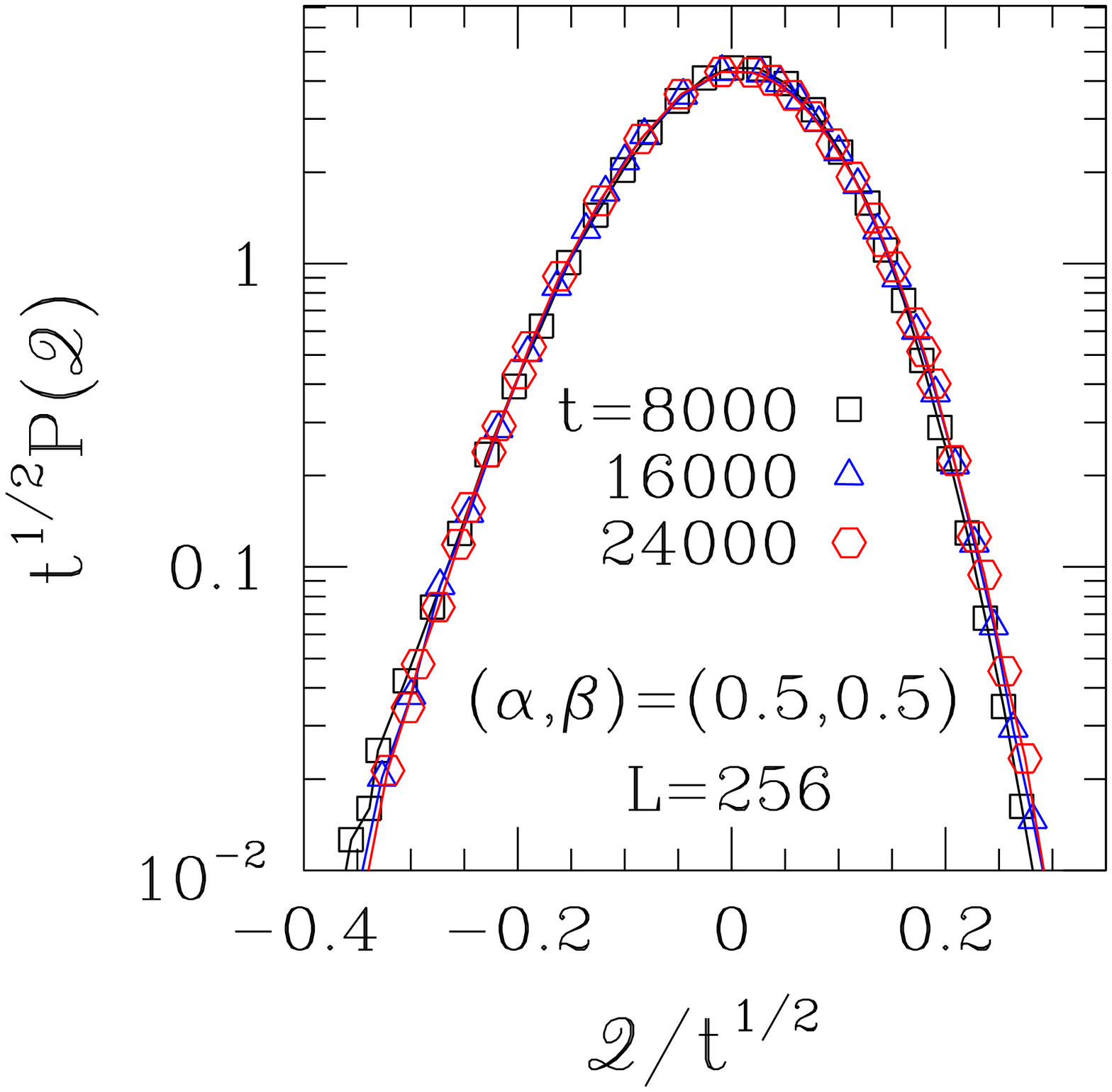}}}
\caption{(Color online) 
Scaling plot of PDF for the variable ${\mathscr Q}(t)$ defined below 
Eq.~(\protect{\ref{eq:jint}}), at $(\alpha,\beta)=(0.5,0.5)$.
$N_{\rm sam}=10^6$ steady-state samples were collected at the indicated times.
} 
\label{fig:qbarab05}
\end{figure}

\begin{figure}
{\centering \resizebox*{3.3in}{!}{\includegraphics*{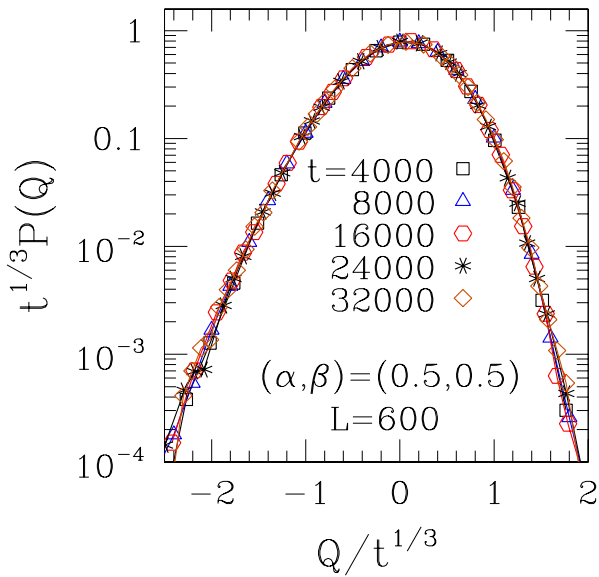}}}
\caption{(Color online) 
Scaling plot of PDF for the variable $Q(t)$ defined 
in Eq.~(\protect{\ref{eq:qtilde}}), at $(\alpha,\beta)=(0.5,0.5)$.
$N_{\rm sam}=10^6$ steady-state samples were collected at the indicated times.
} 
\label{fig:qab05}
\end{figure}

Further information can be extracted from the PDFs, or full counting statistics, of the
integrated currents. Data for ${\mathscr Q}(t)$ at $(\alpha,\beta)=0.5$ are
shown in Figure~\ref{fig:qbarab05} as a scaling plot.
Although the effective scaling power is $1/2$, which is in line with the linear dependence
of $\langle\langle {\mathscr Q}^2\rangle\rangle$ against $t$ seen in Fig.~\ref{fig:q2q2bar},
the scaled PDF shows a slight negative skew. For comparison, Figure~\ref{fig:qab05}
shows the corresponding scaling plot for $Q(t)$ within the anomalous scaling window,
where a negative skew is more clearly present; however, the scaling power is now $1/3$,
as already remarked in Ref.~\onlinecite{kvo10}.

For a proper analysis of the scaled data shown in Figures~\ref{fig:qbarab05} 
and~\ref{fig:qab05}, and comparison with theoretical predictions where 
available, one must keep in mind that the former necessarily pertain to finite
$t$ and $L$, while the latter either assume taking the $L, t \to 
\infty$ limit~\cite{ess11,lm11}, or~\cite{dem95} $L$ is kept fixed
and $t \to \infty$. 

Denoting by $k$ the scaling powers referred to, it turns out
from the scaled variables $u \equiv {\mathscr Q}/t^k$
[$\,Q/t^k\,$] and $P(u) \equiv t^k\,P({\mathscr Q})$ [$\,t^k\,P(Q)\,$]  that
the cumulants of the global [$\,$local$\,$] current are given by
${\mathcal E}_n\ [\,E_n\,]= t^{nk-1}\,\langle\langle u^n\rangle\rangle$~.

So, for the global current ($k=1/2$), one can have both 
$\langle\langle u^2\rangle\rangle$ and ${\mathcal E}_2$ finite for $t \to \infty$, 
while  $\langle\langle u^3\rangle\rangle$
must vanish in the same limit, since ${\mathcal E}_3$ must not diverge. 
By separately analyzing PDF data used in Figure~\ref{fig:qbarab05} for various times 
$4000 \leq t \leq 24000$
we found that, as $t$ increases, ${\mathcal E}_2=\langle\langle u^2\rangle\rangle$
increases from $0.0077$ to $0.0081$, while $\langle\langle u^3\rangle\rangle$
goes from $-2.7\times10^{-4}$ to $-1.4\times10^{-4}$. The "diffusion constant"
${\mathcal E}_2$ can be compared with $E_2  \approx (16\pi L)^{-1/2}=0.0088 \dots$
for $L=256$~\cite{dem95}. Although there appears to be no {\em a priori} reason to assume
equality of these quantities, the closeness of their values is remarkable.
As regards $\langle\langle u^3\rangle\rangle$, its evolution can be approximately matched
by a $t^{-1/2}$ dependence, which would be compatible with a nonzero limiting value
for ${\mathcal E}_3$.

For the local current ($k=1/3$), if $\langle\langle u^2\rangle\rangle \neq 0$ then
$E_2=\langle\langle Q^2\rangle\rangle/t \sim t^{-1/3}$ which is in line with
existing results~\cite{dem95,ess11,lm11}, recalling that at $(\alpha,\beta)=(0.5,0.5)$:
 (i) $E_2(L) \propto L^{-1/2}$~\cite{dem95}, and (ii) the appropriate 
finite-size scaling combination  is $t/L^{3/2}$. Also, 
$E_3$ should approach a finite value if $\langle\langle u^3\rangle\rangle$ is finite.
From the full set of data in Figure~\ref{fig:qbarab05}, we find $\langle\langle u^3\rangle\rangle
=-0.034(1)$~. Separate analysis of subsets of data for different times shows small fluctuations
around the average just quoted, with no apparent up- or downward trend upon increasing $t$.
Comparison of this estimate to the analytic prediction for the thermodynamic limit, 
$E_3=-1/8$~\cite{ess11,lm11} shows that,
though correct in sign, it is still only about $1/4$ of  the expected value.

\begin{figure}
{\centering \resizebox*{3.3in}{!}{\includegraphics*{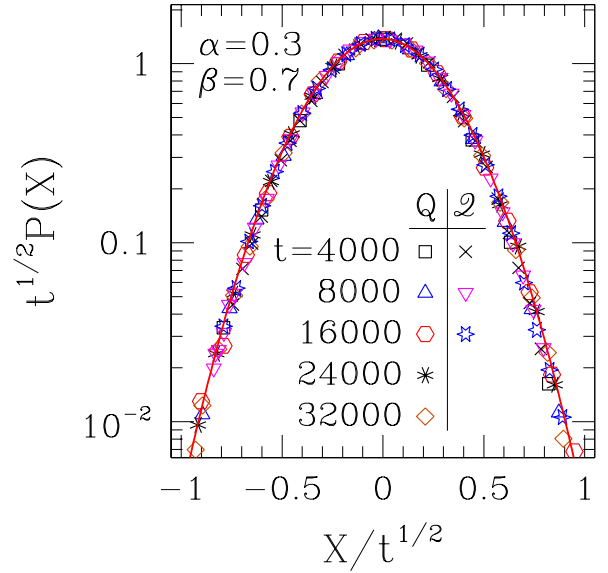}}}
\caption{(Color online) 
Scaling plot of PDFs for: (i) the variable $X=Q(t)$ defined 
in Eq.~(\protect{\ref{eq:qtilde}}) [$\,$ system size $L=600\,$], and (ii)
$X={\mathscr Q}(t)$, defined below  Eq.~(\protect{\ref{eq:jint}})  
[$\,L=256\,$], at $(\alpha,\beta)=(0.3,0.7)$. The full (red) line is a Gaussian fit to
data. $N_{\rm sam}=5\times10^5$ steady-state samples were collected at the 
indicated times~.
} 
\label{fig:qqbara3b7}
\end{figure}

At $(\alpha,\beta)=(0.3,0.7)$, as evinced in Figure~\ref{fig:q2q2bar},
the scaling power is $k=1/2$ for both $\mathscr Q$  and $Q$. Figure~\ref{fig:qqbara3b7}
shows that the scaled PDFs for both quantities are nearly indistinguishable, even when one
plots $L=256$ data for $P({\mathscr Q})$ together with $L=600$ data for $P(Q)$.
%%%%%%%%%%%%%%%%%% Referee's 2nd report, point (3) %%%%%%%%%%%%%%%%%%%
Thus, for this range of $L$, deep inside the LC phase the statistics of
$Q$ and $\mathscr Q$ are essentially identical, and free from finite-size
effects.
%%%%%%%%%%%%%%%%%%%%%%%%%%%%%%%%%%%%%%%%%%%%%%%%%%%%%%%%%%%%%%%%%%%%%%%%%
From fits of pure Gaussians
to the scaled PDFs, we get ${\mathcal E}_2 \approx E_2=0.0841(5)$, in very good 
agreement with the thermodynamic-limit prediction~\cite{dem95,ess11,lm11} $E_2=\alpha(1-\alpha)
(1-2\alpha)=0.084$. Allowing for non-zero skewness gives $E_2 \approx 
0.082$, with $E_3<0$, and $|E_3| \lesssim 1\times 10^{-3}$. However, the estimates for
the latter quantity exhibit uncertainties of order $60\%$ or thereabouts, so they should
be considered with caution.  The theoretical prediction for 
$L \to \infty$~\cite{ess11,lm11} is 
$E_3=\alpha(1-\alpha)(1-6\alpha+6\alpha^2)=-0.0546$~. As in the 
case of ($\alpha,\beta)=(0.5,0.5)$, we get the right sign for $E_3$ but a smaller
absolute value than predicted.

\subsection{$\alpha=\beta=1$}
\label{sec:ab1}
Within the maximal current phase $\alpha,\beta \geq 1/2$, the behavior is expected to
be similar to that observed at $(\alpha,\beta)=(0.5,0.5)$, and reported in 
Section~\ref{sec:a+b1}. At $\alpha=\beta=1$ several simplifications 
occur~\cite{derr93,dem95,lm11,rbsdq12}, allowing for simple expressions to be derived
for the $\langle\langle Q^n \rangle\rangle$, so we ran simulations at that point.

\begin{figure}
{\centering \resizebox*{3.3in}{!}{\includegraphics*{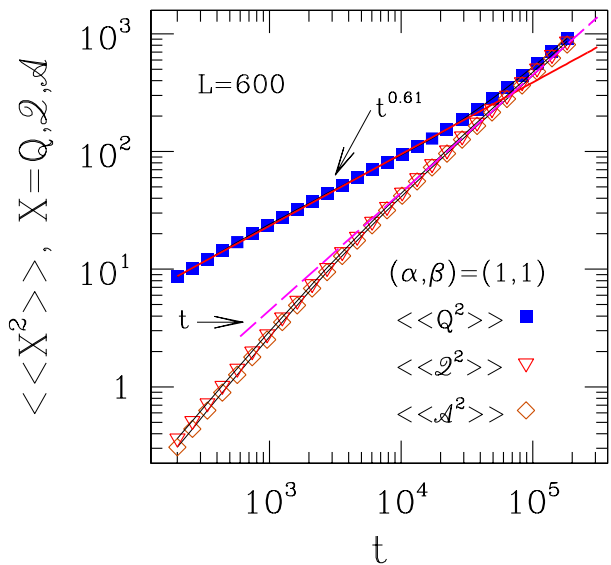}}}
\caption{(Color online) 
At $\alpha=\beta=1$: double-logarithmic plot of
variances $\langle\langle {Q}^2\rangle\rangle$, for
$Q(t)$ defined in Eq.~(\protect{\ref{eq:qtilde}}), 
$\langle\langle {\mathscr Q}^{\,2}\rangle\rangle$ for ${\mathscr Q}(t)$, 
and $\langle\langle {\mathcal A}^{\,2}\rangle\rangle$ for the integrated 
activity ${\mathcal A}(t)$, the latter two
defined below Eq.~(\protect{\ref{eq:jint}}), against time. 
The full line is $\propto t^{0.61}$; the dashed line is $\propto t$.
} 
\label{fig:aq2qbarab1}
\end{figure}
Figure~\ref{fig:aq2qbarab1} shows the time evolution of the variance of the several
integrated quantities under investigation here. Comparing with Figures~\ref{fig:a2qbar}
and~\ref{fig:q2q2bar}, which refer to the same system size and time interval,
the main difference to the behavior exhibited at
$(\alpha,\beta)=(0.5,0.5)$ is in the anomalous scaling of $\langle\langle 
Q^{\,2}\rangle\rangle$. Our best fit to single-power behavior, encompassing 
the same time "window" as used in that case, i.e.,
$200 < t < 3\times 10^4$, gives an exponent equal to $0.61(1)$, close to 
but slightly below the expected value, $2/3$. 
The expected crossover to linear behavior, and merging with the $\langle\langle {\mathscr
Q}^{\,2}\rangle\rangle$ curve, takes place in the same (narrow) time interval as at
$(\alpha,\beta)=(0.5,0.5)$.

\begin{figure}
{\centering \resizebox*{3.3in}{!}{\includegraphics*{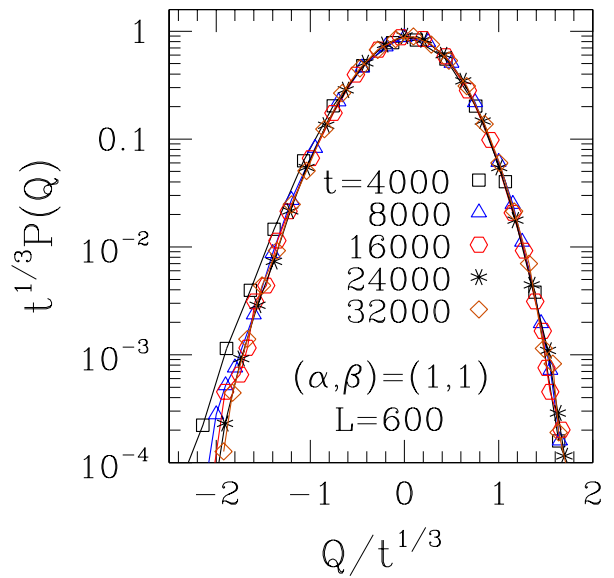}}}
\caption{(Color online) 
Scaling plot of PDF for the variable $Q(t)$ defined 
in Eq.~(\protect{\ref{eq:qtilde}}), at $(\alpha,\beta)=(1,1)$.
$N_{\rm sam}=5 \times 10^5$ steady-state samples were collected at 
the indicated times.
} 
\label{fig:qab1}
\end{figure}

The apparent discrepancy in the exponent value can be solved 
by examining the scaling plot of the
PDF for $Q$, shown in Figure~\ref{fig:qab1}, where the scaling power 
is set to $1/3$. It can be seen that data collapse is rather good,
except for the $t=4000$ data which visibly stray off, especially at the
low end of the curve. Indeed, fitting only data for $8 \times 10^3
 < t < 3\times 10^4$ in Figure~\ref{fig:aq2qbarab1} gives an 
exponent $0.66(2)$. So it is the extent of
the $t^{\,2/3}$ scaling window which is shorter here than for
$(\alpha,\beta)=(0.5,0.5)$~.
Further analysis of the scaling plot depicted in Figure~\ref{fig:qab1} follows the same 
lines as that of Figure~\ref{fig:qab05}: (i) again, $E_2(t) \sim t^{-1/3}$, with the same
interpretation as above; and (ii) the value of $\langle\langle u^3\rangle\rangle =
-0.012(1)$ compares reasonably well to the analytic prediction, $\lim_{L\to \infty} 
E_3=-0.009\dots$~\cite{lm11}. 

\begin{figure}
{\centering \resizebox*{3.3in}{!}{\includegraphics*{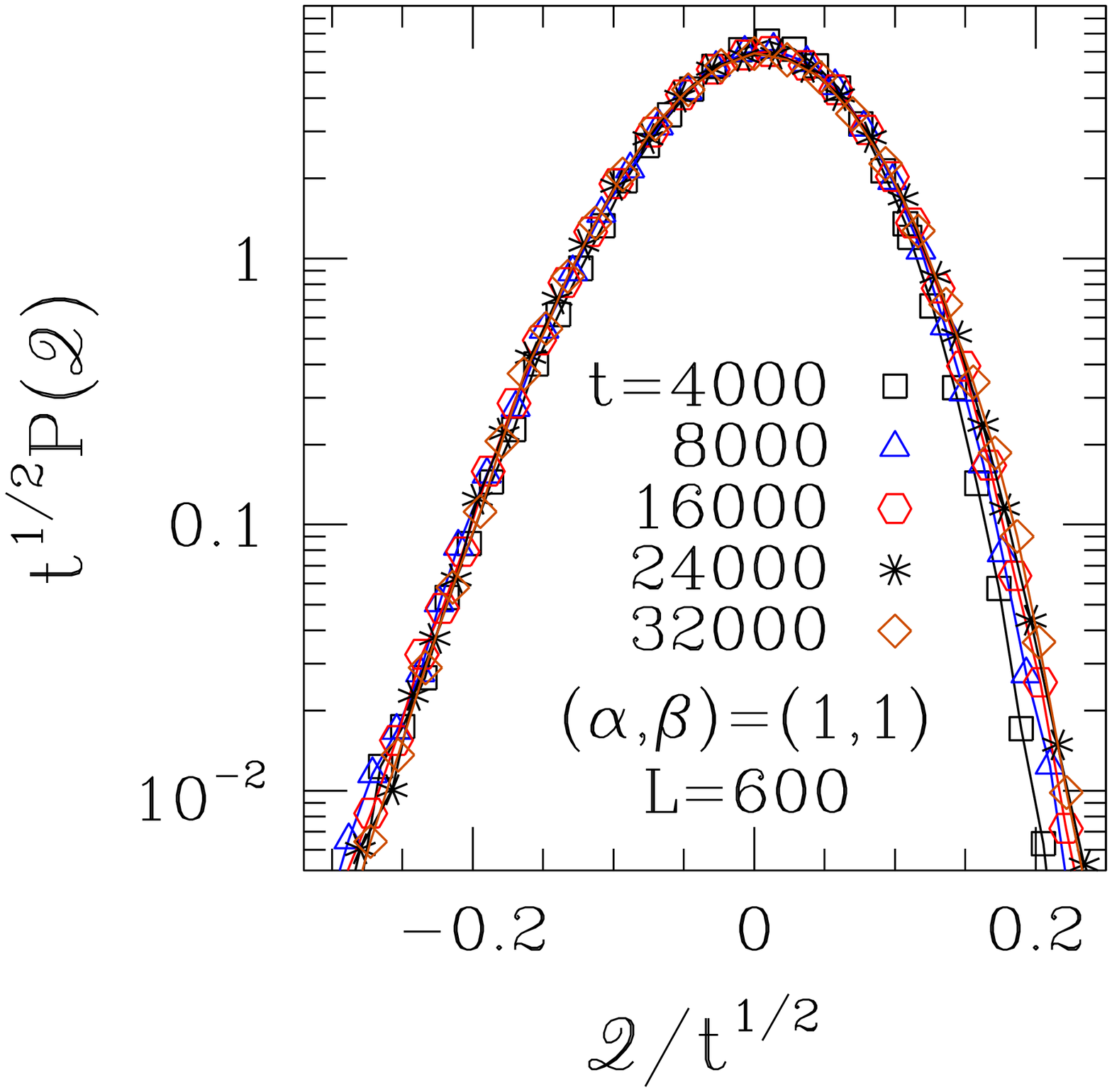}}}
\caption{(Color online) 
Scaling plot of PDF for the variable ${\mathscr Q}(t)$ defined below 
Eq.~(\protect{\ref{eq:jint}}), at $(\alpha,\beta)=(1,1)$.
$N_{\rm sam}=10^6$ steady-state samples were collected at the indicated times.
} 
\label{fig:qbarab1}
\end{figure}
Figure~\ref{fig:qbarab1} shows the scaling plot for the PDF of $\mathscr Q$, using $1/2$ as the scaling 
power. The quality of data collapse is not as good as for points at the
border, or outside, of the maximal-current phase (see, respectively, Figures~\ref{fig:qbarab05}
and~\ref{fig:qqbara3b7}). However, it is possible to find a rather stable estimate for the 
"diffusion  constant", ${\mathcal E}_2=0.0045(1)$. Again, this compares favorably with 
$E_2=(3\sqrt{2\pi}/64)\,L^{-1/2}=0.004797\dots$ for  $L=600$~\cite{lm11}.
The third cumulant is negative but very close to zero,
${\mathcal E}_3 \approx -1\times 10^{-4}$. 

\section{Discussion and Conclusions} 
\label{sec:conc}

We have set out to compare the features exhibited by fluctuations of activity with those
of (local) current for the TASEP with open BCs.

We define $C_n \equiv \langle\langle A^n\rangle\rangle$, with the systemwide (instantaneous) 
activity $A$ defined in Eq.~\ref{eq:adef}, for compatibility with the notation of 
Ref.~\onlinecite{rbsdq12}. So ${\mathcal C}_n \equiv C_n/(L+1)^n$ is its normalized
counterpart.
As shown in Ref.~\onlinecite{rbsdq12}, ${\mathcal C}_2$ vanishes with system size as $L^{-1}$
for any $(\alpha,\beta)$
(see Table~\ref{t1}). 
The standard skewness $S$, defined as 
$S=C_3/C_2^{3/2}={\mathcal C}_3/{\mathcal C}_2^{3/2}$~\cite{numrec}, vanishes as 
$L^{-x}$ (so ${\mathcal C}_3 \sim L^{-(x+3/2)}$), usually with $x=1/2$ ($x=3/2$) in the low 
(high)-current phase, except for some special lines or points on the phase diagram   
(see Figure~\ref{fig:pd}). 
As remarked in Section~\ref{sec:th}, ${\mathcal C}_2$ and ${\mathcal C}_3$ vanish with 
increasing $L$ 
because they reflect fluctuations of the equal-time, nonlocal correlations 
of ${\mathcal O}(L)$ stochastic variables.

On the other hand, cumulants of the integrated activity ${\mathcal A}$, as well as those
of the position-averaged current ${\mathcal J}$, behave similarly (albeit with different
decay times on the approach to steady state, see Figures~\ref{fig:q2q2bar} 
and~\ref{fig:aq2qbarab1}) to those of the local current, except that they do not capture the 
anomalous scaling which occurs 
for $\alpha,\beta \geq 1/2$. This indicates that the systemwide sampling at equal times
overshadows the subtler effects which come about from non-equal time correlations,
within the maximal-current phase and along its border. Anomalous scaling 
can be unveiled 
only when the local current is considered, in which case systemwide, 
equal-time sampling is absent, and the system-size dependence is much less relevant
overall 
(although it still shows up in the width of the scaling "window", 
see Ref.~\onlinecite{kvo10} and Figures~\ref{fig:q2q2bar} and~\ref{fig:aq2qbarab1} above).

Going back to the skewness of distributions, the sign of $S$ is positive for ${\rm 
min}\,(\alpha,\beta) \leq \frac{1}{2}-\frac{\sqrt{5}}{10}=0.27639 \dots$, 
and negative otherwise, with the following
exceptions: $(\alpha,\beta)=(0.5,0.5)$ where $S \equiv 0$, and $(\alpha,\beta)=(1,1)$,
$(1,0.5)$, and $(0.5,1)$ where $S >0$~\cite{rbsdq12}. 
At  ${\rm min}\,(\alpha,\beta)=0.27639 \dots$, $S>0$ but depends on $L^{-3/2}$, as opposed
to the $L^{-1/2}$ form which holds generally in the low-current phase.
All such exceptions are accidental,
i.e., they occur because of symmetries or cancellations which are valid only for those
specific values of $(\alpha,\beta)$~\cite{rbsdq12}. 
Furthermore, it has been shown that $S>0$ is
to be expected on physical grounds for $\alpha \ll 0.5$ in the low-current, low-density phase (see
Figure 3 of Ref.~\onlinecite{rbsdq12} and accompanying remarks). 

Concerning local-current statistics, 
we recall the results given in Table~\ref{t2}:
for the low-current phase Refs.~\onlinecite{ess11,lm11} predict that, with $\gamma \equiv
{\rm min}\,(\alpha,\beta)$, $E_3=\gamma(1-\gamma)(1-6\gamma+6\gamma^2)$ which is thus
zero for $\gamma_0=\frac{1}{2}-\frac{\sqrt{3}}{6}=0.2113 \dots$~, and positive 
(negative) for $\gamma < \gamma_0$ ($\gamma > \gamma_0$).
Ref.~\onlinecite{lm11} evaluates $E_3 <0$ at $(\alpha,\beta)=(1,1)$, as recalled in
Section~\ref{sec:ab1}, and gives arguments showing that the dominant behavior of the
cumulants should be the same as that at $(\alpha,\beta)=(1,1)$ everywhere within the 
maximal-current phase.

In conclusion, once accidental exceptions are understood as such, 
both $S$ for activity and $E_3$ for the local current tell essentially
the same story: fluctuation distributions have positive skew for low injection
(or ejection) rates, and negative skew otherwise. The borderline between the
two types of behavior lies deep within the low-current phase, and its relationship
to the actual (second-order) transition to the maximal-current phase it at best that 
of a precursor.
 
As a final remark, we emphasize that accurate numerical checks of the
theoretical predictions for $E_3$ would need much longer simulations than those
reported here, since evaluation of this quantity strongly depends on a proper description of 
the tails of the local-current PDF. However, at the level of accuracy pursued here, 
the present results fulfil our goal of providing a comparative
analysis of the main features of activity- and local-current fluctuations.

\begin{acknowledgments}
The author thanks the Brazilian agencies  
CNPq  (Grant No. 302924/2009-4) and FAPERJ (Grant No. E-26/101.572/2010)
for financial support.
\end{acknowledgments}


\begin{thebibliography}{99}

\bibitem{derr98} B. Derrida, Phys. Rep. {\bf 301}, 65 (1998).
\bibitem{sch00} G. M. Sch\"utz, in {\it Phase Transitions and
Critical Phenomena}, edited by C. Domb and J. L. Lebowitz (Academic,
New York, 2000), Vol. 19.
\bibitem{derr93} B. Derrida, M. Evans, V. Hakim, and V. Pasquier, 
J. Phys. A {\bf 26}, 1493 (1993).
\bibitem{rbs01} R. B. Stinchcombe, Adv. Phys. {\bf 50}, 431 (2001).
\bibitem{be07} R. A. Blythe and M. R. Evans, J. Phys. A {\bf 40}, R333 (2007).
\bibitem{cmz11} T. Chou, K. Mallick, and R. K. P. Zia, Rep. Prog. Phys. 
{\bf 74}, 116601 (2011).
\bibitem{dem93} B. Derrida, M. R. Evans, and D. Mukamel, J. Phys. A {\bf 26}, 4911 (1993).
\bibitem{dem95} B. Derrida, M. R. Evans, and K. Mallick, J. Stat. Phys. {\bf 79}, 833 (1995).
\bibitem{dl98} B. Derrida and J. L. Lebowitz, \prl {\bf 80}, 209 (1998). 
\bibitem{bd06} T. Bodineau and B. Derrida, J. Stat. Phys. {\bf 123}, 277 (2006).
\bibitem{kvo10} T. Karzig and F. von Oppen, \prb {\bf 81}, 045317 (2010).
\bibitem{gv11} M. Gorissen and C. Vanderzande, J. Phys. A {\bf 44}, 115005 (2011).
\bibitem{lm11} A. Lazarescu and K. Mallick, J. Phys. A {\bf 44}, 315001 (2011).
\bibitem{ess11} J. de Gier and F. H. L. Essler, \prl {\bf 107}, 010602 (2011).
\bibitem{rbsdq12}  R. B. Stinchcombe and  S. L. A. de Queiroz, \pre {\bf 85}, 041111 
(2012).
\bibitem{ds04} M. Depken and R. Stinchcombe, \prl {\bf 93}, 040602 (2004).
\bibitem{ds05} M. Depken and R. Stinchcombe, \pre {\bf 71}, 036120 (2005).
\bibitem{rsss98}  N. Rajewsky, L. Santen, A. Schadschneider, and
M. Schreckenberg, J. Stat. Phys. {\bf 92}, 151 (1998).
\bibitem{dqrbs08} S. L. A. de Queiroz and  R. B. Stinchcombe, 
\pre {\bf 78}, 031106 (2008).
\bibitem{nas02} Z. Nagy, C. Appert, and L. Santen,  J. Stat. Phys. {\bf 109},
623 (2002). 
\bibitem{ess05} J. de Gier and F. H. L. Essler, \prl {\bf 95}, 240601
(2005); J. Stat. Mech.: Theory Exp. (2006) P12011.
\bibitem{numrec} W. Press, B. Flannery, S. Teukolsky, and W. Vetterling,
{\it Numerical Recipes in Fortran, The Art of Scientific
Computing}, 2nd ed. (Cambridge University Press, Cambridge, England, 1992),
Chap 14. 
\end{thebibliography}
\end{document}